\documentstyle[11pt,aps,preprint,epsf,floats]{revtex}

\preprint{UCRHEP-T203}

\def\hp{H^+}
\def\ho{H^0}
\def\hpm{H^\pm}
\def\mh{m_{\hp}}
\def\lesim{\,{\raise-3pt\hbox{$\sim$}}\!\!\!\!\!{\raise2pt\hbox{$<$}}\,}
\def\gesim{\,{\raise-3pt\hbox{$\sim$}}\!\!\!\!\!{\raise2pt\hbox{$>$}}\,}

\def\whatjournal{P}
\newlinechar=`\^^J

\def\ordernpb#1#2#3{{\bf#1} (#3) #2}

\if P\whatjournal {\global\def\order#1#2#3{\orderprd{#1}{#2}{#3}}}
                    \immediate\write16{^^J PRD references ^^J}\else 
                   {\global\def\order#1#2#3{\ordernpb{#1}{#2}{#3}}}
                    \immediate\write16{^^J NPB references ^^J} 
\fi

\def\mpla#1#2#3{{\rm  Mod. Phys. Lett. {\bf A}}\order{#1}{#2}{#3}}
\def\npb#1#2#3{{\rm Nucl. Phys. {\bf B}}\order{#1}{#2}{#3}}
\def\plb#1#2#3{{\rm Phys. Lett. {\bf B}}\order{#1}{#2}{#3}}
\def\prl#1#2#3{{\rm Phys. Rev. Lett.\ }\order{#1}{#2}{#3}}
\def\prd#1#2#3{{\rm Phys. Rev. {\bf D}}\order{#1}{#2}{#3}}
\def\zphys#1#2#3{{\rm Z. Phys. {\bf C}}\order{#1}{#2}{#3}}

\begin{document}

\title{Enhanced Three-Body Decay of the Charged Higgs Boson }
\author{ Ernest Ma$^1$, D.P Roy$^{1,2}$, and Jos\'{e} Wudka$^1$}
\address{$^1$Department of Physics, University of California. % \\
Riverside, California, 92521-0413\\
%.\\
$^2$Tata Institute of Fundamental Research, Mumbai 400 005, India}

\date{\today}

\maketitle

\begin{abstract}
If the charged Higgs boson $\hp$ exists with $\mh < m_t + m_b$, the 
conventional expectation is that it will decay dominantly into $c \bar s$ 
and $\tau^+ \nu_\tau$.  However, the three-body 
decay mode $\hp \rightarrow W^+ b \bar b$ is also present and we
show that it becomes very important in the low 
$\tan \beta$ region for $\mh \gesim 140$ GeV.  We then explore its phenomenological 
implications for the charged-Higgs-boson search in top-quark decay.
\end{abstract}

\bigskip\bigskip
%\newpage

The discovery of the top quark at the Tevatron collider~\cite{cdf.d0,tipton}
has generated a good deal of current interest in the search for new
particles in the decay of the top-quark. In particular, top quark decay
is known to be a promising reaction to look for the charged Higgs boson of a
two-scalar doublet model and, in particular, the minimal supersymmetric
standard model (MSSM)~\cite{mssm}. In the diagonal CKM matrix approximation
the MSSM charged Higgs boson couplings to the fermions are given by
\begin{equation}
{\cal L} = { g \over \sqrt{2} \; m_W } \hp \left[ 
\cot \beta \; m_{u \, i} \bar u_i d_i{}_L + 
\tan \beta \; m_{d \, i} \bar u_i d_i{}_R +
\tan \beta \; m_{\ell \, i} \bar \nu_i \ell_i{}_R \right] + \mbox{h.c.}
\label{lagr}
\end{equation}
where $ \tan \beta $ is the ratio of the vacuum expectation values of
the two scalar doublets and the index $i$ labels the quark and lepton
generation. This interaction implies a large $ \hp t b $ Yukawa coupling
when
\begin{equation}
\tan \beta \lesim 1 \quad \mbox{and} \quad \tan \beta \gesim m_t / m_b 
\end{equation}
where one expects a large branching fraction for $ t \rightarrow b \hp
$ decay (given $ m_t > \mh$). Interestingly the regions $ \tan \beta
\sim 1$ and $ \gesim m_t/m_b $ are favored by SUSY-GUT models for a
related reason -- {\it i.e.} the unification of the $b$ and $ \tau $
masses which requires a large negative contribution from the top
Yukawa coupling to the renormalization group equation~\cite{rg}.

It should be noted here that the perturbation theory limit on the $ \hp
t b $ Yukawa coupling requires
\begin{equation}
0.2 < \tan \beta < 100
\end{equation}
while the GUT scale unification constraint implies stricter limits
\begin{equation}
1 \le \tan \beta \le m_t / m_b
\end{equation}
which are also required if one assumes the perturbation theory limit on
the Yukawa coupling to remain valid up to the GUT scale~\cite{pert}.
Without any GUT scale Ansatz, however, the allowed region of $ \tan
\beta $ extends down to $0.2 $. We shall assume only the particle
content of the MSSM Higgs sector but no constraints from GUT scale
physics. Our analysis will remain valid in any two-Higgs doublet model
satisfying the coupling pattern of the MSSM as given by (\ref{lagr});
{\it i.e.} the so-called class II models~\cite{class.2}.

For $ m_t > \mh $ the dominant decay modes are usually assumed
to be the two body decays $ \hp \rightarrow c \bar s , \ \tau^+ \nu $.
The corresponding widths are
\begin{eqnarray}
\Gamma_{cs} &=& { 3 g^2 \mh \over 32 \pi m_W^2 } \left( m_c^2 \cot^2
\beta + m_s^2 \tan^2 \beta\right ) \label{gam.cs} \\
\Gamma_{\tau \nu} &=& { g^2 \mh \over 32 \pi m_W^2 } 
m_\tau^2 \tan^2 \beta \label{gam.taunu}
\end{eqnarray}
The leading QCD correction is taken into account by substituting the
quark mass parameters for eqs. (\ref{lagr}) and (\ref{gam.cs}) by the
running masses at the $\hp$ mass scale. Its most important effect is to
reduce the charm quark mass $ m_c $ from $1.5 $GeV to $1$GeV~\cite{mc}.
Consequently the two rates are approximately equal when $ \tan \beta
\sim 1 $; the $ \tau \nu $ ($ c s $) rate dominates when $ \tan \beta >
1$ ($ \tan \beta < 1 $).

In this note we shall consider the phenomenological implications of
a very important 3-body decay channel of the Higgs boson, namely
\begin{equation}
\hp \rightarrow \bar b b W^+ ,
\label{3.bod.decay} 
\end{equation}
where the $b W^+$ comes from a virtual $t$ quark~\cite{3bod}. The dominant
contribution comes from the top-quark exchange with a large Yukawa
coupling of $ \hp $ to the top quark
given by the first term in eq. (\ref{lagr}). One can
easily calculate the corresponding width as
\begin{equation}
{ d \Gamma_{ \bar b b W } \over d s_{\bar b } d s_b } = { 1 \over 256
\pi^3 \mh^3 } \left( { 3 g^4 m_t^4 \cot^2 \beta \over 4 m_W^4 \left(
m_t^2 - s_{\bar b } \right)^2 } \right) \left[ m_W^2 \left( s_W -2 m_b^2
\right) + \left( s_{ \bar b }- m_b^2 - m_W^2 \right) \left( s_b - m_b^2
- m_W^2 \right) \right] 
\end{equation}
where $s_{\bar b } $, $s_b $ and $ s_W $ are the 4-momentum squared
transferred to the corresponding particles satisfying $ s_{\bar b } + s_b 
+  s_W = \mh{}^2 + m_W^2 + 2 m_b^2 $~\cite{mand}.

\begin{figure}[ht]
\centerline{ \vbox to 3.5 in {\epsfxsize=6 truein\epsfbox[0 -200 612 592]{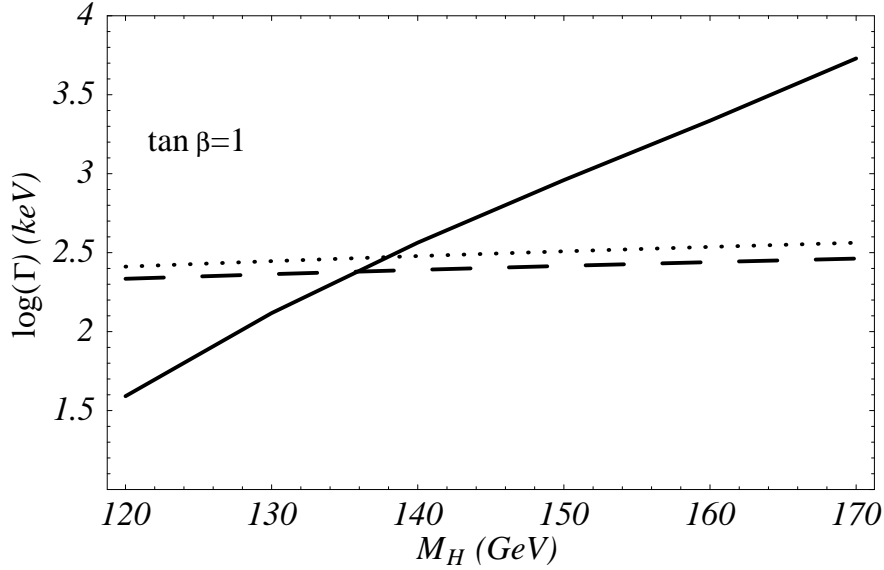}}}
\caption{Comparison of the 3 body decay width $ \Gamma_{ \hp
\rightarrow \bar b b W } $ (solid)  with the 2 body widths $ \Gamma_{
\hp \rightarrow c \bar s } $ (dashed) and $ \Gamma_{ \hp \rightarrow 
\tau^+ \nu } $ (dots).
\label{fig:f1}}
\end{figure}

Figure~\ref{fig:f1} compares the 3 body decay width $ \Gamma_{ \bar b b
W } $ with the 2 body widths $ \Gamma_{ cs } $ and $ \Gamma_{ \tau \nu }
$ over the charged Higgs boson mass range $ 120 - 170 $GeV at $ \tan
\beta =1 $. $ \Gamma_{ \bar b b W } $ is seen to be the dominant decay
width for $ \mh \gesim 140 $GeV, while the 2 body decays dominate up to
$ \mh = 130 $GeV. The reason for this is the large $\hp $ Yukawa
coupling to $ t \bar b $, which is about 100 time larger than those to
the $ c \bar s $ and $ \tau^+ \nu $ channels. This can overcome the
extra suppression factors due to the gauge coupling of the $W$ as well
as the 3 body phase space, provided the off-shell propagator suppression
factor is not too large. The latter is ensured for $ \mh \gesim 140
$GeV. Thus the 3 body decay (\ref{3.bod.decay}) is the dominant mode for
\begin{equation}
\mh \gesim 140 \mbox{GeV} \quad \mbox{and} \quad \tan \beta \lesim 1 
\end{equation}
while the $ \tau \nu $ mode (\ref{gam.taunu}) dominates at larger $ \tan
\beta $. The $ c \bar s $ mode is relatively small at all $ \tan \beta $
for $ \mh \gesim 140 $GeV. It may be noted here that the relative size
of the $ \hp $ decay widths at $ \tan \beta = 1 $ (Fig.~\ref{fig:f1})
would hold for all values of $ \tan \beta $ in the two-Higgs doublet
model of type I~\cite{class.2}.

This situation has a close parallel in the neutral scalar sector. For a
neutral Higgs $ \ho $ whose mass is slightly below the $WW$ threshold a
good detection channel is $ WW^*$ with $ W^* \rightarrow \ell \nu $. In
this case the decay $ \ho \rightarrow W \ell \nu $ is comparable to $ \ho
\rightarrow \bar b b $~\cite{class.2}. A related decay $ \hp
\rightarrow W^+ Z^* $ with $ Z^* \rightarrow b \bar b $ is not
considered because for multi-doublet models there is no $ \hp W^- Z $
coupling~\cite{hp.w.z}.

The $ \hpm $ search strategies in top quark decay have so far been based
on the distinctive features of the channels
\begin{eqnarray}
&& t \rightarrow b \hp \rightarrow b \tau^+ \nu \label{t.taunu} \\
&& t \rightarrow b \hp \rightarrow b c \bar s \label{t.cs}
\end{eqnarray}
vis-a-vis the standard model decay
\begin{equation}
t \rightarrow b W^+ \rightarrow b \; ( \ell \nu , \, \tau \nu , \, q'
\bar q ) \label{t.bw}
\end{equation}
As we have seen above, however, this strategy is valid only up to $ \mh
\simeq 130 $GeV. For $ \mh \gesim 140 $GeV the $ c \bar s $ mode
(\ref{t.cs}) is overtaken by
\begin{equation}
t \rightarrow b \hp \rightarrow b \bar b b W^+ \rightarrow b \bar b b (
l \nu, \; \tau \nu, \;  q' \bar q ) \label{t.bbbw}
\end{equation}
as the dominant decay mode for the low $ \tan \beta $ ($ \lesim 1$)
region. The distinctive feature of this new channel is evidently very
different from those of the channels (\ref{t.taunu}) and ( \ref{t.cs}).

In order to assess the impact of the new channel (\ref{t.bbbw}) let us
summarize the main features of the current $ \hp $ search program in $ t
\bar t $ decay. It is based on two strategies --- {\it i)} excess of $
t \bar t$ events in the $ \tau $ channel, and {\it ii)} their deficit in
the leptonic ($ \ell = e , \, \mu $) channel with respect to the
standard model prediction from (\ref{t.bw}). The first is appropriate
for the large $ \tan \beta $ region where the $ \tau \nu $ channel
(\ref{t.taunu}) is the dominant channel of the charged Higgs decay. One
can already get significant limits on $ \mh $ for very large $ \tan
\beta $ ($ \gesim m_t / m_b $) from the CDF $ t \bar t $ data in the $
\ell \tau $ and inclusive $ \tau $ channels~\cite{cdf.ltau,cdf.t.incl}.
This analysis can be extended down to lower values of $ \tan \beta $ at
the Tevatron upgrade and the LHC by exploiting the opposite states of $
\tau $ polarization from $ W^\pm $ and $ \hpm $ decays~\cite{lhc}.
Evidently this type of analysis would not be affected by the new
channel.

The second strategy is based on a suppression of the leptonic ($e , \,
\mu $) decay of the top due to the $ \hp $ channels (\ref{t.taunu}) and
(\ref{t.cs})~\footnote{This is evident for the $ cs $ channel
(\ref{t.cs}) but should also hold for the $ \tau \nu $ channel
(\ref{t.taunu}) as well since the $ e , \, \mu $ from $ \tau $ decay are
expected to be soft and hence suppressed by the $ p_T$ cut used in the
analysis.}. The experimental estimate of the $ t \bar t $ cross section
is based on the $ \ell \ell $ and $ \ell + $multijet channels with a $
b-$ tag, requiring leptonic decay of at least one of the top quarks. Thus
the presence of the $ \hp $ channels (\ref{t.taunu}) and (\ref{t.cs})
would imply a decrease of this $ t \bar t$ cross section, while the
experimental estimate~\cite{d0.cdf.ttbar}
\begin{equation}
\sigma_{ t \bar t } ( \mbox{CDF+D\O} ) = 6.5 {\renewcommand{\arraystretch}{.3}
\scriptsize  \begin{array}{l} +1.3 \\
-1.2 \end{array}}  \mbox{pb} \qquad
\sigma_{ t \bar t } ( \mbox{CDF} ) = 7.6 {\renewcommand{\arraystretch}{.3}
\scriptsize  \begin{array}{l} +1.8 \\
-1.5 \end{array} } \mbox{pb} 
\end{equation}
is actually slightly higher than the QCD prediction of $ \sigma_{ t
\bar t } \le 5.6 $pb~\cite{qcd.ttbar}. This has lead to a significant
lower limit on $ \mh $ at low $ \tan \beta $ ($ \lesim 1 $), assuming
dominance of the $ cs $ decay channel
(\ref{t.cs})~\cite{mh.bound1,mh.bound2}. Evidently this method will be
valid only up to $ \mh = 130 $GeV. Beyond this value the dominant charged Higgs
decay channel in the low $ \tan \beta $ ($ \lesim 1 $) region is
(\ref{t.bbbw}), which does not imply any reduction in the leptonic decay
of the top. Instead it implies an increase in the $b-$tagging efficiency
due to the multi$-b$ final state. Since the CDF cross section is largely
based on the $b-$tagged events, the presence of the decay channel
(\ref{t.bbbw}) would imply an increase of this cross section relative
to the standard model prediction, instead of a decrease. Thus it will
go in the same direction as the data.

\begin{figure}[ht]
\centerline{ \vbox to 3.5 in {\epsfxsize=6 truein\epsfbox[0 -200 612 592]{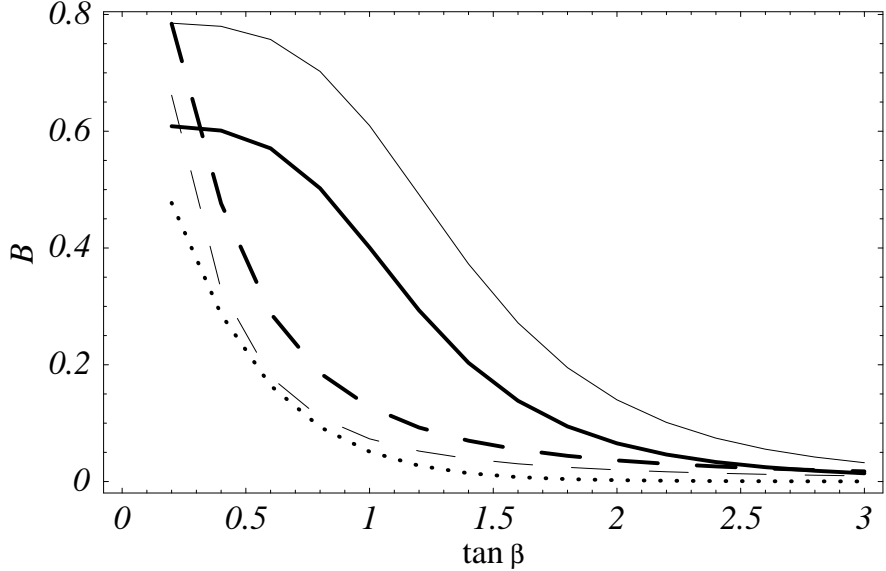}}}
\caption{Branching fractions for $ t \rightarrow b \hp $ (dashed lines)
and $ \hp \rightarrow \bar b b W $ (solid lines) decays for low $ \tan
\beta $. Heavy lines and thin lines correspond to $ \mh = 140 $ and $
\mh = 150 $GeV respectively. The dotted line corresponds to the product
$ B( t \rightarrow b \hp ) B(\hp \rightarrow \bar b b W )$ for $ \mh =
140$GeV (the plot for $ \mh = 150 $GeV is practically identical).
\label{fig:f2}}
\end{figure}

Let us now look at the implications of the new $ \hpm $ decay channel
(\ref{t.bbbw}) on Tevatron $ t \bar t $ events more closely. In
Fig.~\ref{fig:f2} we show the branching fractions for $ t \rightarrow b
\hp $ and $ \hp \rightarrow \bar b b W $ decays over the low $ \tan
\beta $ region for $ \mh = 140 $ and $ 150 $GeV. Also shown is the
product of these two branching fractions,
\begin{equation}
B = B ( t \rightarrow b \bar b b W ) = B ( t \rightarrow b \hp ) B ( \hp
\rightarrow \bar b b W )
\end{equation}
which is about the same for both values of $ \mh $. We see that this
branching fraction lies in the range $ 5 - 20\% $ for $ \tan\beta =
1-0.6 $. This corresponds to a probability of about $ 10-40\% $ ($ \simeq
2B $) for the channel
\begin{equation}
\bar t t \rightarrow \bar b b \bar b b W W 
\label{tt.bbbbww}
\end{equation}
where one of the top-quarks decays via an $ \hpm $ and we have
made a first-order approximation in $B$. Thus the $2b$ and $4b$ final
states occur with relative probabilities $1-2B $ and $2B$ respectively,
where the former also includes a small contribution from the 2-body
decays of the $ \hpm $.. 

It should be mentioned here that the decay of the $ \hpm $ into a
neutral Higgs and a real or virtual $W$ boson is (whenever kinematically
allowed) an additional source for a $4b$ final state such as (\ref{tt.bbbbww}).
Within the MSSM this contribution can be significant over the low $ \tan \beta
$ region~\cite{3bod} depending on the SUSY breaking parameters. 
Thus the 3-body decay considered above constitutes
a minimal contribution to the $4b$ final state (\ref{tt.bbbbww})
generated by the decays of the charged Higgs boson.

We have studied the characteristic features of the above channel versus
the standard model decay
\begin{equation}
\bar t t \rightarrow \bar b b W W 
\label{ttbar.bbww}
\end{equation}
via a parton-level Monte Carlo program. While the $\ell $ and $ \nu $
from $W$ decay have very similar kinematic distributions in the two cases,
there is a clear difference in the number of tagable $b$-quarks. The CDF
SVX detector has a tagging efficiency of $ \epsilon_b = 0.24 $ per $b$ 
satisfying
\begin{equation}
E_T^b > 20 \mbox{GeV}, \quad | \eta_b | < 2 
\label{b.tag}
\end{equation}
which takes into account the loss of efficiency due to the limited
rapidity coverage of the vertex detector ($ | \eta_{\mbox{\tiny SVX}} | \lesim 1
$)~\cite{epsb.run2}. This is expected to go up to  $ \epsilon_b = 0.4 $ per
$b$ for run II as the rapidity coverage of the vertex detector is
extended to $ | \eta_{\mbox{\tiny SVX}} | = 2 $. Table~1 shows the
probability distribution of the numbers of $b$ quarks per event
satisfying the tagging criterion (\ref{b.tag}) for the signal
(\ref{tt.bbbbww}) and the standard model background
(\ref{ttbar.bbww}) channels. It shows that the majority of the signal
events are expected to contain 3-4 tagable $b$ quarks for $ \mh = 140
$GeV (similar results hold for $ \mh = 150$GeV). It also shows the
probability distribution for the expected number of $b$-tags per event
for the SVX tagging efficiency of $ \epsilon_b = 0.24 $, where we have
assumed  that the uncorrelated probability for tagging $n$ out of $N$
tagable $b$-quarks is $ P_n^N = { N \choose n } \epsilon_b^n ( 1 - 
\epsilon_b )^{N - n} $. The
corresponding expectations for the run II efficiency $ \epsilon_b = 0.4
$ are shown in parenthesis. The implications for the $ t \bar t $ events in
the $b$-tagged $ \ell+$multijet channel are discussed below.

\bigskip\bigskip

\begin{tabular}{|c||c|c|c|c||c|c|c|}
\hline
 \multicolumn{1}{|c||}{\hspace*{3pt}Probability (\%)\hspace*{3pt}}
&\multicolumn{4}{c||}{No. of tagable $b$'s/event}
&\multicolumn{3}{c|}{No. of $b$-tags/event}\\
& 1 & 2 & 3 & 4 & $\ge$1 & $\ge$2 & $\ge$3 \\
\hline
$\bar t t \rightarrow \bar b b \bar b b W W $ $(2B)$ &\hspace*{6pt}  4.7\hspace*{6pt} &\hspace*{6pt}  25.6 \hspace*{6pt}  & \hspace*{6pt} 50.6 \hspace*{6pt} &\hspace*{6pt} 18.9 \hspace*{6pt}& \hspace*{3pt}52.8 (74.2) \hspace*{3pt}& \hspace*{3pt}12.4 (31.8)\hspace*{3pt} & \hspace*{3pt}(6.6) \hspace*{3pt}\\
\hline
$\bar t t \rightarrow \bar b b W W $ $ ( 1 - 2B ) $  & 13  & 87   &  --  &  --  & 39.6 (60.9) &    5 (13.4) &   --  \\
\hline
\multicolumn{8}{l}{Table 1. \parbox[t]{5in}{\small\baselineskip 10 pt Probabilities for 
different numbers of tagable $b$ quarks per
event and numbers of $b$-tags (per event) with $ \epsilon_b = 0.24 \;
(0.4) $ for the $ \hpm $ signal ($ \mh = 140$GeV) and the standard
model background.}}
\end{tabular}

\bigskip\bigskip

As we see from this table
the probability of inclusive single ($ \ge1 $) $b$-tag
is $52.8\%$ for the signal compared to $ 39.6\%$ for the standard model
decay, {\it i.e.}, about $1/3$ higher. 
Consequently the measured $ \bar
t t$ cross section will appear larger than the standard model prediction
by $ (1/3) \times (2 B) $, 
{\it i.e.} about $13\%$ for $ B = 0.2 $. This could
account for at least part of the excess of the CDF $ t \bar t$ cross
section~\cite{d0.cdf.ttbar} over the standard model prediction. Even more
significantly, the probability for inclusive double ($ \ge 2$) $b$-tag is
$ 12.4 \% $ for the signal compared to only $5\%$ for the standard model
decay, {\it i.e.} an excess of 150\%. This would imply an excess of double
$b$ tagged events over the standard model prediction by $ 3 B $, {\it i.e.} 
$ 60\% $ for $ B =0.2 $. Again there seems to be an indication of
such an excess in the CDF data~\cite{new.into.b}. It should be remarked
however that the excess is expected to appear in the $ \ge 3 $ jet
events; but not in the $2$ jet sample, except through fluctuations. 
It is therefore premature to link the reported excess to the above 
mechanism. It is important to note, however, that the size
of the signal can have visible impact even at the level of the
existing limited data.

It should be noted here that one expects a 20-fold rise in the number of
$ t \bar t$ events in the run II, and the efficiency of single and
double tags to go up by factor of $1.5$ and $ \sim 3$ respectively. Thus
one expects about $1000$ single ($\ge1$) and $200$ double ($\ge2$)
$b$-tagged events for CDF, and similar numbers for D\O\ (in run II). Even
with a $B$ of only $5\%$, this would correspond to an excess of $
\sim30$ double $b$-tagged events, {\it i.e.} a $ 2-3 \sigma $ effect.
Moreover, the $6.6\%$ efficiency for $ \ge3$ $b$-tags for the signal would
imply at least $10-12$ triple $b$-tagged events for $ B \ge 5\%$.
Finally, one should be able to get additional constraints from the
clustering of the reconstructed $ \hpm $ mass.

Thus the 3-body decay channel provides a visible signature for a charged
Higgs boson in top-quark decay over its region of dominance, {\it i.e.}
$ \mh \gesim 140 $GeV and $ \tan \beta \lesim 1 $. This can be used to
probe for an $ \hpm $ at the Tevatron run II over the mass range $
140-150$GeV, and can be extended beyond $160$GeV at the LHC. We conclude
with the hope that this channel will play an important role in the
charged Higgs boson search program in the future. 

\acknowledgments
We thank Prof. V. Barger for discussions and Dr. M. Mangano
and Dr. G.P.
Yeh for several communications regarding the CDF $b$ tagging efficiency. This
work was supported in part by the Department of Energy under Grant
DE-FG03-94ER40837.

\end{document}